\documentclass[apj]{emulateapj}
\def\arcsec{\hbox{$^{\prime\prime}$}}

\def\cm2{cm$^{-2}$}
\def\cc{cm$^{-3}$}

\def\nh3{NH$_3$}
\def\n2h{N$_2$H$^+$}

\def\co{$^{12}$CO}
\def\13co{$^{13}$CO}
\def\c18o{C$^{18}$O}
\def\hc3n{HC$_3$N}
\def\h2{H$_2$}
\def\nh{n(H$_2$)}

\def\lc{\>\> ,}
\def\Ms{M$_{\odot}$}

\begin{document}

\title{Is the Taurus B213 Region a True Filament?:  Observations of Multiple Cyanoacetylene Transitions}
\author{Di Li\altaffilmark{1,2,3}, Paul F. Goldsmith\altaffilmark{3}}

\altaffiltext{1} {National Astronomical Observatories, Chinese Academy of Sciences,  Chaoyang District, Datun Rd, A20, Beijing 100012, China, \em{email:}
dili@nao.ac.cn }
\altaffiltext{2} {Space Science Institute, Boulder, CO, USA}
\altaffiltext{3} {Jet Propulsion Laboratory, California Institute of Technology, Pasadena, CA, USA}
\begin{abstract}

We have obtained spectra of the J=2-1 and J=10-9 transitions of
cyanoacetylene (\hc3n) toward a collection of positions in the most
prominent filament, B213, in the Taurus molecular cloud. The
analysis of the excitation conditions of these transitions reveals
an average gas \h2\ volume density of $(1.8\pm 0.7 ) \times10^{4} $
\cc. Based on column density derived from 2MASS and this volume
density, the line of sight dimension of the high density portion of
B213 is  found to be $\simeq$ 0.12 pc, which is comparable to the
smaller projected dimension and much smaller than the elongated
dimension of B213 ($\sim$2.4 pc). B213 is thus likely a true
cylinder--like filament rather than a sheet seen edge-on. The line
width and velocity gradient seen in \hc3n are also consistent with
Taurus B213 being a self-gravitating filament in the early stage of
either fragmentation and/or collapse.

\end{abstract}

\section{Introduction}
 The advent of telescopes with large format imaging arrays, such as Herschel and JCMT, makes possible maps of star
 forming regions with large spatial dynamic ranges.  It has become apparent in such images in both gas (Narayanan et al.\  2008) and
 dust emission (Men'shchikov et al.\ 2010; Arzoumanian et al.\ 2011) that, when properly sampled, molecular gas clouds exhibit filamentary
 structures on many spatial scales. This confirms the morphology that had been identified earlier  in studies of visual extinction (Schneider \& Elmegreen 1979).
  The origin of such elongated structures is unclear. A filament as a projected two dimensional structure, can be consistent with different theories.
  In particular, models suggest that gravitationally bound cigar--like clouds will collapse to become spindles, while pancake-like clouds will continue
  to flatten under the influence of their self-gravity to become sheet-like structures (Lin et al.\ 1965; McKee \& Ostriker 2007).
  In shock-induced molecular cloud formation models, the region around shock front will also be flattened (V{\'a}zquez-Semadeni et al.\ 2006).

\begin{figure*}[htp]
\includegraphics[width=0.7\textwidth, angle=-90]{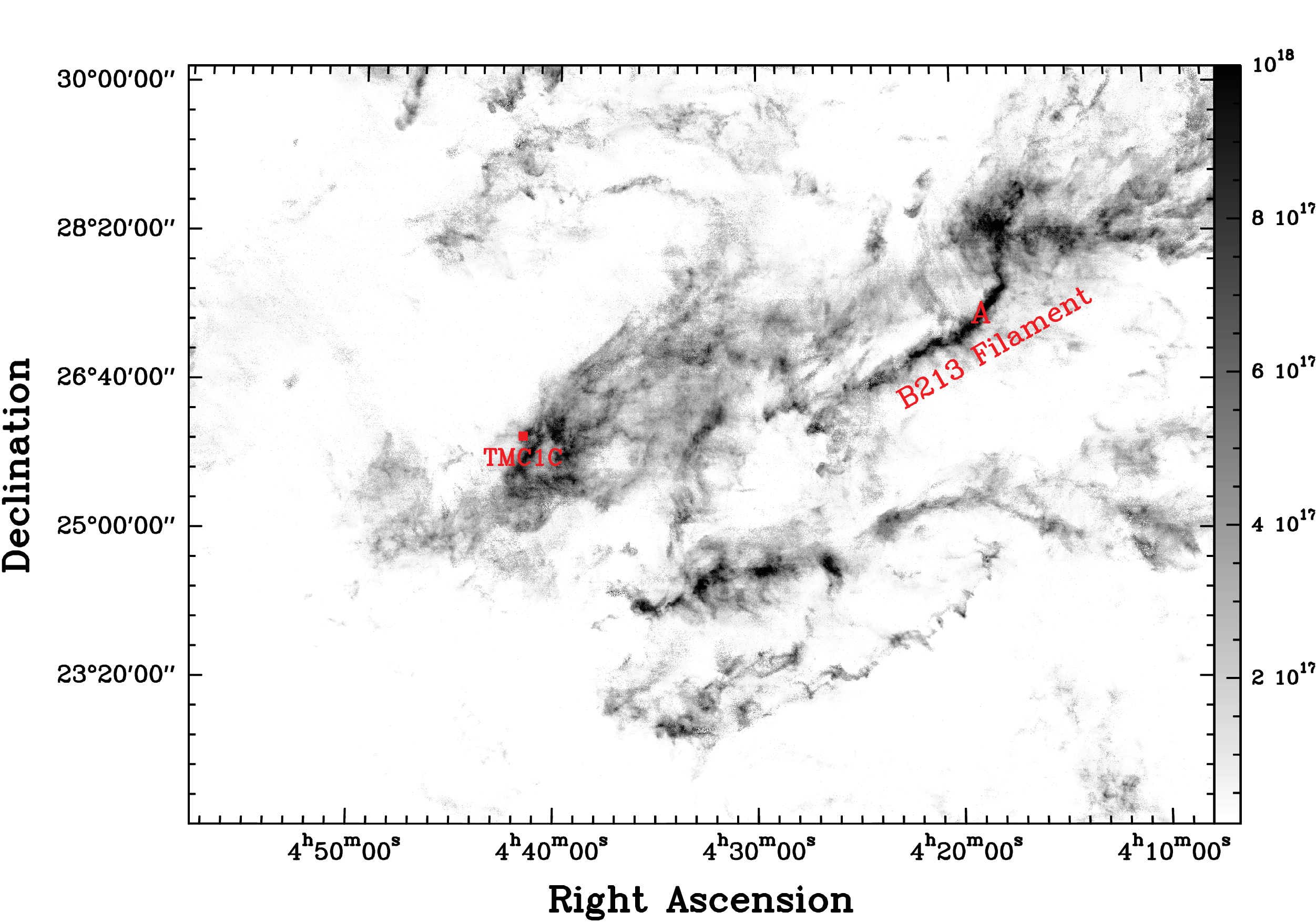}
\caption{ \co\ column density map of the Taurus molecular cloud
containing B213 filament. The coordinates in this and subsequent
figures are J(2000). The red letter "A" indicates the location of
\hc3n\ measurements and is the same as the position A in Fig.~
\ref{obs}. The red square indicates TMC1C. \label{b213}}
\end{figure*}

To distinguish a true filamentary (cylinder like) structure from a
sheet seen projected essentially edge-on, we have sought to obtain
the line of sight dimension  of a prominent dense filament in
Taurus, B213 (Fig.~\ref{b213}). Taurus is one of the closest star
forming regions. The portion of Taurus molecular cloud that contains
B213 filament has been covered by a rich collection of surveys at
optical through radio wavelengths (Goldsmith et al.\ 2008; Rebull et
al. 2010).  For this study, we have obtained data along ``cuts" in
the direction orthogonal to the elongation of the filament, in order
to evaluate the variation of the excitation conditions in the
filament, and to determine its volume density from which the line of
sight dimension can be determined.

The column density of the molecular gas can be determined by
optically thin tracers including gas and dust. Goldsmith et al.\
(2008) presented a column density map of the whole Taurus region
through analysis combining \13co 1-0 and \co\ 1-0 transitions. There
have been only a few previous attempts to determine the volume
density of gas in Taurus. The first of these was by Avery et al.\
(1982), who observed a few positions towards TMC-1, a well-known
high-density region. Avery et al. (1982) obtained five transitions
(J=4-3, 5-4, 9-8, 10-9, and 11-10) of cyanoacetylene (HC$_3$N) at
millimeter wavelengths. Their excitation analysis  found a halo
component having density 8$\times$10$^3$ \cc\ and a core having a
density 6$\times$10$^4$ \cc. TMC-1 is part of the observed ring-like
structure, but is not really a filament as are many other structures
seen in Taurus.
 Schloerb, Snell \& Young (1983) studied the same region in three transitions of \hc3n, and found a single density ~$10^5$ \cc\  to fit their data satisfactorily.
 If the $\sim$1 to 2 arcminute telescope beams are filled by gas at these relatively high densities, the line of sight dimension must be quite small,
 about 0.1pc, in order to be consistent with the column density.
 Considering the molecular ring to be a tube of circular form, the diameter of the tube is no more than 0.4 pc.
 This does not give clear evidence for any extreme geometry.
 Pratap et al.\  (1997) derived the density for positions in TMC1 based on the J=4-3, 10-9, and 12-11 transitions of \hc3n.
 The range of densities are generally consistent with the core value from Avery et al.\ (1992), but Pratap et al.  did not observe TMC1C.
 Onishi et al. (2002) identified a large number of cores in Taurus using H$^{13}$CO$^+$ J = 1-0.
 The detection of this spectral line is generally suggestive of high densities, ~10$^5$ \cc, but since only a single transition of this
 species was observed, the density could not be directly determined.
 Stepnik et al. (2003) made a cut across a dense filament in Taurus at multiple wavelengths between 200 to 400 microns.
 Their model (which has large uncertainties) suggests a peak density $\sim$ 6$\times$10$^4$ \cc, which suggests a true filamentary geometry.

In this study, we focus on using J=2-1 and J=10-9 transitions of \hc3n as a density probe, which has the following advantages.
The excitation of these spectral lines is sensitive to the volume density in likely range of this quantity of relevance to Taurus.
When the 2-1 transition at 18.2 GHz is observed with the 100 meter Green Bank Telescope (GBT), the resulting FWHM beam size is close to 40\arcsec,
which is similar to the 69\arcsec\ beam of the 12 meter telescope of the Arizona Radio Observatory (ARO) at the 90.979 GHz frequency of the 10-9 transition.

\section{Observations}
\begin{figure}[htp]
\includegraphics[width=8cm, angle=-90]{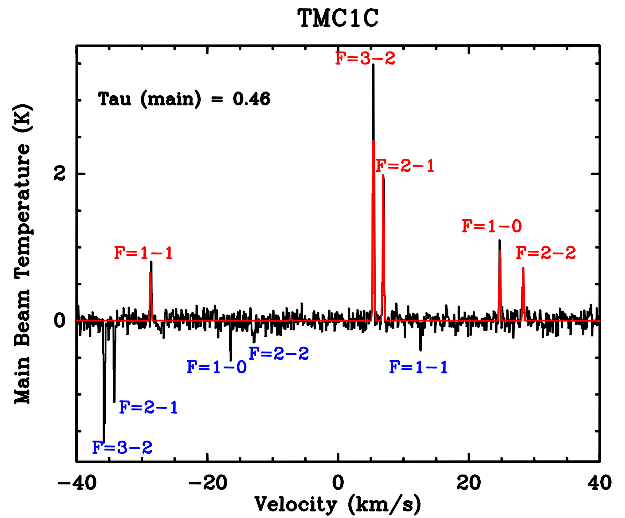}
\caption{Spectrum of the \hc3n J=2-1 line toward TMC1C. A
double-frequency-switching mode was employed to obtain these data.
The negative peaks (blue labels) are a consequence of the folding
process used to analyze a frequency switched spectrum. The negative
F=2-2, F=1-0, F=2-1, and F=3-2 components are a result of a shift to
higher frequency, while the F=1-1 negative component is due to a
shift to lower frequency. The intensity scale is the main beam
antenna temperature, as described in the text. The hyperfine fit to
the spectrum gives peak antenna temperature (corrected for optical
depth) of the F=3-2 component to be 5.3 K, V$_{lsr}$ equal to 5.4
km/s, and FWHM equal to 0.21 km/s. \label{tmc1c-fit}}

\end{figure}

    At the Green Bank Telescope (GBT), we obtained data on the \hc3n J=2-1 transition toward positions across several filaments and
    TMC1C in the Taurus molecular cloud.
The data were taken in two runs on 12 and 25 February 2009. The
observations were carried out in frequency switching mode with a
frequency throw of 3 MHz, which is relatively small compared to the
total bandwidth of 50 MHz. The total number of channels is 8096 with
a velocity resolution of 0.1 km/s. The resulting spectra have flat
baselines and the emission is easily recovered through folding
procedures (Fig.~\ref{tmc1c-fit}). The intensity calibration was
done following the procedure in the GBTidl manual written by J.
Braatz. The calibration requires the knowledge of atmospheric zenith
optical depth at 18 GHz, which was provided by the GBT weather
forecast tool to be 0.0183 for 12 February and 0.035 for 25
February. The resulting antenna temperatures were further scaled to
main beam temperatures by using a main beam efficiency of 0.88
(provided by GBTidl).

\begin{figure}[htp]
\includegraphics[width=8cm, angle=-90]{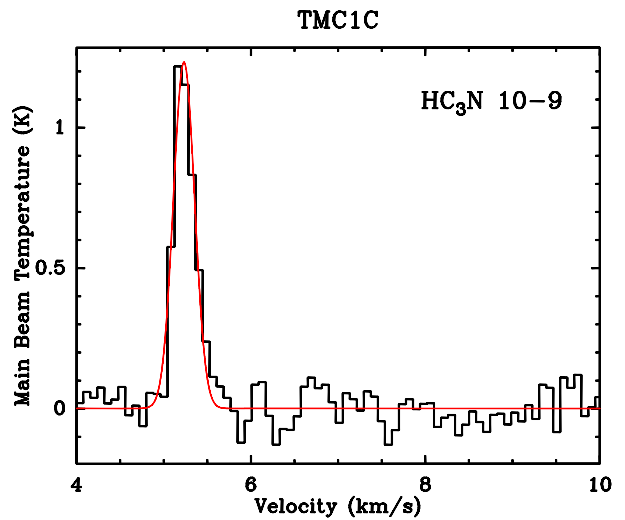}
\caption{Spectrum of the \hc3n J=10-9 line toward TMC1C. A single
Gaussian fit gives V$_{lsr}$ equal to 5.2 km/s, FWHM equal to 0.29
km/s, and integrated area equal to 0.38 K km/s. The peak of the
fitted Gaussian is 1.23 K. \label{tmc1c-10-9}}
\end{figure}

At the Arizona Radio Observatory 12m telescope (ARO), we observed
the \hc3n 10-9 and 12-11 transitions toward the same set of
filaments. Only the 10-9 transition was detected and only in
positions across the B213 filament. Between 2009 and 2010, we also
carried out three runs to map out the spatial extent of the 10-9
emission in B213.  The \hc3n\ data were taken in position switching
mode at ARO. A simple correction was then applied to scale the data
to main beam antenna temperature using the main beam efficiency of
0.87 at 91 GHz from the ARO online manual. As seen in
Fig.~\ref{tmc1c-10-9}, the hyperfine components of the 10-9
transitions are not resolved.

\section{Results of Data Analysis}
    In all observing runs at both GBT and ARO, we observed TMC1C regularly to check the system and use it as a secondary flux calibrator.
    The analysis of these spectra are discussed in the following subsection using TMC1C as an example.
    The scientific analysis in \S\ref{density} and \S\ref{3D} will focus on observations of B213.

\subsection{TMC1C}
TMC1C exhibits strong \hc3n\ J=2-1 emission with peak main beam antenna temperature of the F=3-2 component being over 3 K.
We did not detect the 2-1 line toward other targets in Taurus except for the B213 filament.
This is  a reflection of a much lower abundance of \hc3n compared to that of TMC1C.
The chemical differentiation in the Taurus region, although an interesting topic by itself, is beyond the scope of this paper.
The dense core TMC1C has been known to exhibit strong emission in transitions of carbon chain molecules.
We observed this source both as a relative intensity calibrator and as a test of our analysis procedures.

The \hc3n 2-1 transition has 6 hyperfine components, five of which can be clearly seen in our spectrum of TMC1C.
We fit all hyperfine components using the``hfs" fitting method in the GILDAS CLASS package (http://www.iram.fr/IRAMFR/GILDAS).
The basic assumptions are a single line width for all hyperfine components and no blending, both of which are reasonable for TMC1C.
The fitted line characteristics are given in Table 2. The fitted peak optical depth is 0.47.
This is a modest opacity, estimated with the reasonable assumption that excitation temperature of all hyperfine components are the same.

The optical depth of the F=3-2 hyperfine component, and the corrected peak antenna temperature of the F=3-2 component are given in columns 7 and 5 of Table 2.
The integrated intensity of each hyperfine component is corrected for its own optical depth, and the corrected integrated intensities for
the J=2-1 transition are given in column 6 of Table 2.

\subsection{The B213 Filament}
\begin{figure}[htp]
\includegraphics[width=8cm, angle=0]{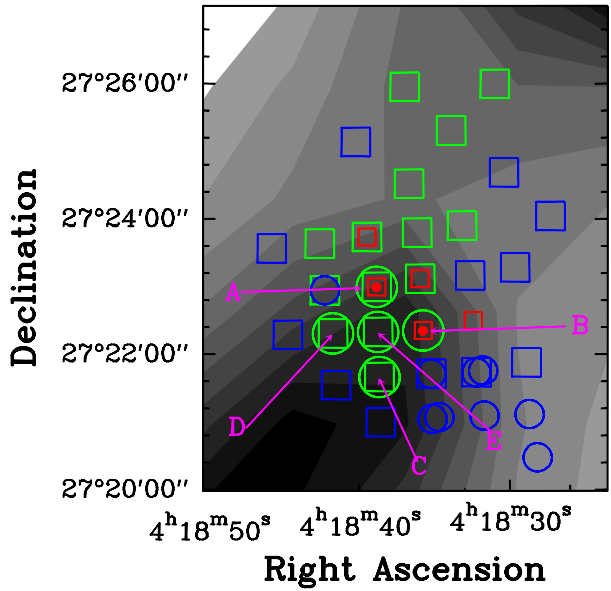}
\caption{A summary of our pointings and \hc3n\ detections. The
underlying image is of magnitude of 2MASS extinction (Pineda et al.\
2010). The blue symbols represent non-detections, the green symbols
represent detection of the 5.6 km/s component, and the red symbols
represent detection the 6.7 km/s component (see text). The circles
denote the J=2-1 transition and the squares denote the J=10-9
transition. The combination of shape and color represents whether
emission is a particular velocity component is detected or not in a
particular transition. \label{obs}}
\end{figure}

The data toward B213 were taken in the same setup as used for TMC1C.
We observed a number of positions, which are indicated in
Fig.~\ref{obs}, which shows all observed lines of sight toward B213
in both the J=2-1 and the J=10-9 transition. We have detected both
10-9 and 2-1 transitions in a total of 5 adjacent positions. The
coordinates for these positions are given in Table 1. The extent of
emission is delineated on three sides by non-detections and is well
constrained within the ridge. There are two velocity components of
\hc3n in this region, which correspond to components seen in \13co.
The main component is at around 5.7 km/s, while the second component
peaks at around 6.8 km/s. In the direction orthogonal to the
elongation of the filament, the size of \hc3n\ emission is between 2
and 3 arcminutes. The analysis of the density structure will focus
on these 5 positions.

\begin{figure}[htp]
\includegraphics[width=8cm, angle=-90]{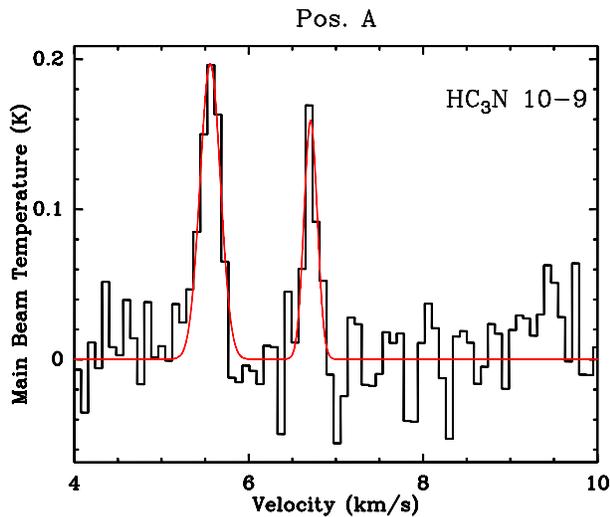}
\caption{Spectrum of the \hc3n\ J=10-9 line toward position A in the
B213 filament. The red line is the result of a two velocity
component Gaussian fit with parameters given in the text and in
Table 2. \label{a_10_9}}
\end{figure}

These spectra are complicated by the fact that there are two
velocity components in this filament. The displacement of these two
components is about 1 km/s, which makes them blended in tracers with
larger velocity dispersion, such as CO and its isotopologues. Due to
the narrowness of \hc3n lines, the two velocity components are
clearly resolved. In Fig.~\ref{a_10_9}, we show the J=10-9 line at
position A. A two component Gaussian fit reveals the peak velocities
to be 5.6 and 6.7 km/s. The peak temperatures are 0.20 and 0.16 K,
and the FWHM 0.28 and 0.20 km/s, respectively.

\begin{figure}[htp]
\includegraphics[width=8cm, angle=-90]{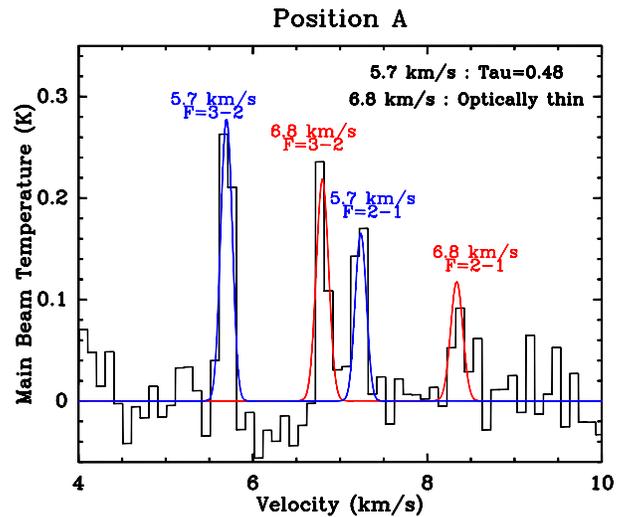}
\caption{Spectrum of the \hc3n J=2-1 line toward position A in the
B213 filament. The velocity scale is referenced to the strongest
(F=3-2) hyperfine component, which shows two velocity components, at
5.7 km/s and 6.8 km/s. The peaks at 7.3 and 8.5 km/s are due to the
F = 2-1 hyperfine component. The blue and red lines represent the
hyperfine fits including two velocity components. \label{a_2_1}}
\end{figure}

The J=2-1 line of \hc3n\ in B213 contains both hyperfine and
multiple velocity components. In Fig.~\ref{a_2_1} we show the J=2-1
spectrum of position A, in which a total of four peaks are visible
due to the two stronger hyperfine components for each of the two
velocity components, on the same velocity range as that of
Fig.~\ref{a_10_9}. The two observed hyperfine components of the 2-1
line are separated by about 1.5 km/s. We have performed hyperfine
fits for each of the velocity components to derive their optical
depths. The lower velocity component is found to have a peak optical
depth of 0.48, similar to that of TMC1C. The higher velocity
component is found to be optically thin. The total column densities
of TMC1C and B213 traced by 2MASS extinction (Pineda et al.\ 2010)
are similar. The difference in line strength by a factor of 10 or
more is thus attributable to a much higher \hc3n\ abundance in
TMC1C. This chemical variation is not a focus of this paper and does
not affect the excitation analysis and the derived volume density in
the following sections.

 \begin{deluxetable}{lcc}
 \tablewidth{0pt}
 \tablecolumns{3}
 \tablecaption{Source List}
\tablehead{
\colhead{Source} & \colhead{Right Ascension }& \colhead{Declination} \\
& \colhead{ (J2000)}& \colhead{(J2000)} }\\
\startdata
TMC1C&04:41:38.82&26:00:42.3\\
B213 A&04:18:38.45&27:23:00.6\\
B213 B&04:18:35.45&27:22:20.6\\
B213 C&04:18:38.45&27:21:40.6\\
B213 D&04:18:41.45&27:22:20.6\\
B213 E&04:18:38.45&27:22:20.6\\
\enddata
\end{deluxetable}

\section{Density Determination through Excitation Analysis of HC$_3$N Transitions}
\label{density}
Filaments can be defined as elongated structures in two spatial dimensions
observed in images of both dust and gas. Such structures are found
to be prevalent in star forming regions. The physics of filament formation and its
implication for star formation hinges on the three dimensional geometry of such
structures. The essentially unknown line of sight (los) dimension of structures that have very unequal dimensions in the plane of the
sky (pos) is critical for determining whether the structures are filaments if d$_{los}$ is comparable to the smaller pos
dimension or sheets viewed edge on if d$_{los }$ is comparable to the larger pos dimension.
To determine d$_{los }$, one need both volume density and column density.
In this section, we derive the volume density based on excitation analysis of \hc3n\ transitions.

In order to compare our data with statistical equilibrium calculations of the excitation of the observed \hc3n lines,
we correct the integrated intensity of the J=2-1 line by the optical depth derived through fitting the hyperfine components.
The integrated intensity of each hyperfine component is corrected for its own optical depth.
The line ratio in Table 2 refers to the ratio between the total J=2-1 integrated intensity and the integrated intensity of the J=10-9 line.
This line ratio is what is modeled  by the intensities given by our radiative transfer calculations  in which the hyperfine components are
not explicitly considered.

The situation for analyzing collisional excitation of cyanoacetylene in order to determine
 the local density has improved significantly of late.  Green and Chapman (1978) carried out close-coupling
calculations of rotational excitation in collisions with helium representing
molecular hydrogen in its lowest, J = 0, spherically symmetric state.  These calculations
were used in some of the works cited above.  The first calculations that treated
 ortho- and para- H$_2$ as distinct species were carried out by Wernli et al. (2007a).
The improvements in the calculation of the potential energy surface were expected to
yield improved accuracy for the collision rates.  However, as explained by Wernli et al. (2007b),
problems in the numerical realization resulted in the results for ortho--H$_2$ being invalid,
while those for para--H$_2$ are thought to be satisfactory.

Quantum calculations by Faure and Wiesenfeld (2011) of collisions between ortho--H$_2$ and HC$_3$N
have recently been made available to us.
Collisions with para--H$_2$ show a propensity for $\Delta J$  = 2 collisions, but the rates
are only modestly smaller for $\Delta$ J = 1 and
$\Delta$ J = 3.
The new rates for collisions with ortho--H$_2$ are quite different, favoring  $\Delta J$ = 1.
The result is that while the largest ortho--H$_2$ rates are up to a factor of 4 larger than the para--H$_2$ rates, the overall
collision rates are more nearly equal.
Both the ortho--H$_2$ and para--H$_2$ rates are modestly larger than the Green and Chapman rates, even when the latter
are scaled by a factor 1.39 to account for the lower reduced mass of H$_2$ compared to that of He.

\begin{figure}[htp]
\includegraphics[width=8cm, angle=0]{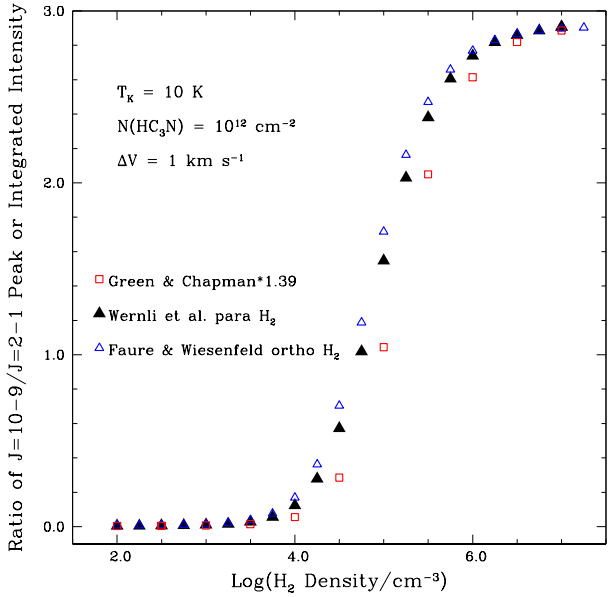}
\caption{Dependence of the ratio of \hc3n\ J = 10-9 to J = 2-1
intensities as a function of H$_2$ density.  The molecular column
density is chosen so that all transitions are optically thin; in
this case the ratios for the peak and the integrated temperatures
are the same. The three sets of points correspond to the Green \&
Chapman (1978) rates, the Wernli et al. (2007a) rates for
para-H$_2$, and the Faure \& Wiesenfeld (2011) rates for
ortho--H$_2$. The observed antenna temperature ratio of 0.4
corresponds to a H$_2$ density $\simeq$ 10$^4$ cm$^{-3}$.
\label{rat-all}}

\end{figure}

We have used the RADEX code (Van der Tak et al.\ 2007) to calculate
the statistical equilibrium level populations and resulting line
intensities from a molecular cloud. We adopt a column density
N(HC$_3$N) = 10$^{12}$ cm$^{-2}$ and a line width of 1 km/s, which
ensures that all transitions are optically thin, and which is close
to $N$(HC$_3$N) determined for the B213 filament, in what follows.
In Fig.~\ref{rat-all} we show the ratio of the intensity of the J =
10-9 to that of the J = 2-1 transition as a function of molecular
hydrogen density, for a kinetic temperature of 10 K. For optically
thin emission and equal line widths (as found in the present data),
the ratios of peak and integrated temperatures are the same. As seen
in Fig.~\ref{rat-all}. the line ratio reaches specified value for a
lower density of ortho--H$_2$ than of para--H$_2$. The difference,
however, is only about a factor of 1.3 to 1.5 in density.

The ortho to para H$_2$ ratio (OPR) is quite uncertain.
If the molecular hydrogen is formed with its internal levels populated at a temperature comparable to the molecule's binding energy,
the OPR should just reflect the relative statistical weights of the two spin modifications, since their energy difference is so small in comparison.
This would result in an OPR equal to 3.
On the other hand,  equilibration at the cloud temperature should result in a ratio
determined by the J=1 to J= 0 energy difference.
Since this is equivalent to 171 K, the result is an OPR close to zero
for a dark cloud kinetic temperature of 10 K -- 15 K. Observational results are
difficult to obtain, and are mixed, but suggest that an OPR $\simeq$ 1 is a reasonable
upper limit even in moderately warm regions (Neufeld, Melnick \& Harwit 1998).
This may be a reflection of timescales for interconversion of the spin modifications.

Due to the very modest difference between the ortho and para collision rates, even though one might have to consider OPR between 0 and 1, the
excitation of \hc3n\ is only very weakly dependent on this poorly--known quantity.
In dark clouds there is evidence for enhanced molecular deuteration,
which requires very low ortho H$_2$ abundance (Flower, Pineau Des For{\^e}ts \& Walmsley 2006).
Given the above considerations, we will adopt the para--H$_2$ rates (i.e. an OPR = 0) for obtaining densities in the B213 filament.

\begin{figure}[htp]
\includegraphics[width=8cm, angle=0]{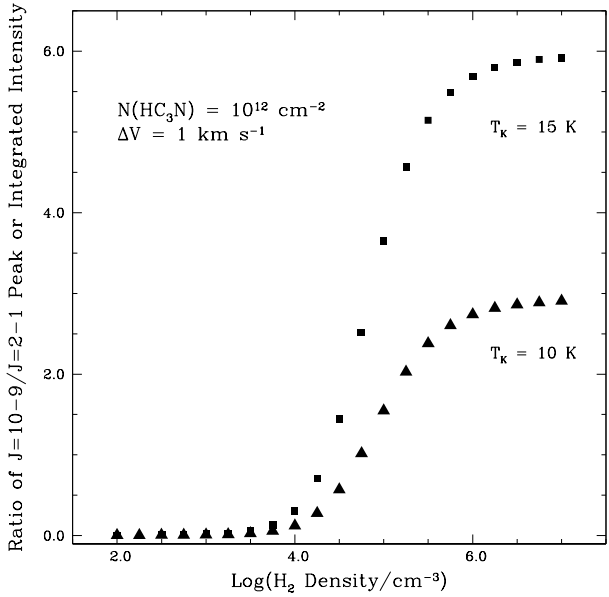}
\caption{Ratio of \hc3n\ J = 10-9 to J = 2-1 transition as a
function of the H$_2$ density, for kinetic temperatures of 10 K and
15 K. The collision rates used are those of Wernli et al. (2007a)
for para-H$_2$ (J = 0).  The reduction in the derived density that
results when the observed intensity ratio is less than 0.5 is
relatively small, less than a factor of 2, in changing the kinetic
temperature from 10 K to 15 K. \label{rat-tk}}

\end{figure}

Our imprecise knowledge of the kinetic temperature results in an additional uncertainty.
\co\  data suggests a kinetic temperature between 12 K and 15 K,
but this may apply more to the outer portion of the filament,
since the lower transitions of this species are very optically thick and
do not really sample the interior of the filament.  It is reasonable
 that the the inner portions in a region such as this with a H$_2$ column
density corresponding to $\simeq$ 10 mag of visual extinction will
be significantly cooler (Evans et al. 2001; Shirley et al. 2005).
Thus, for the optically thin HC$_3$N emission, a kinetic temperature
of 10 K seems appropriate. In Fig.~\ref{rat-tk} we show the effect
of changing the kinetic temperature from 10 K to 15 K. The hydrogen
density inferred for a given observed ratio is lower for the higher
temperature, but in the range of values observed ($\leq$ 0.5), the
density difference does not exceed a factor of 2.

\begin{figure}[htp]
\includegraphics[width=8cm, angle=-90]{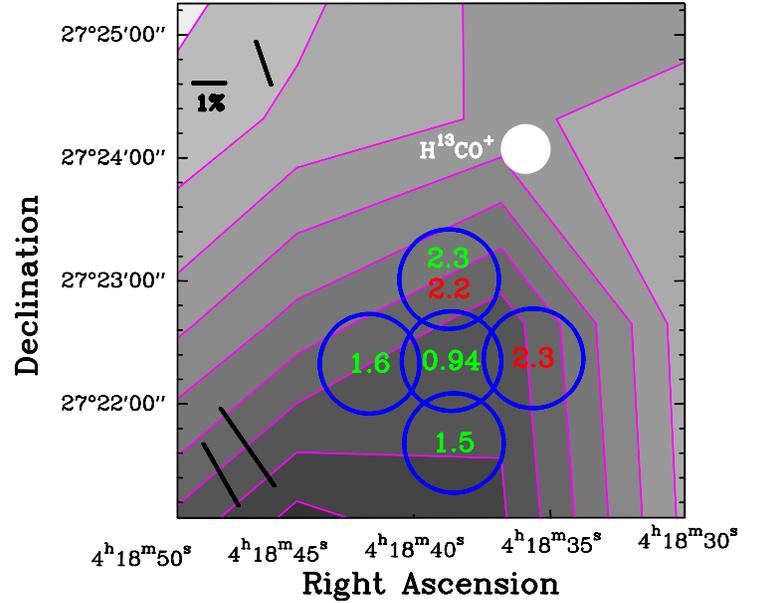}
\caption{The measured volume densities for the 5.6 km/s (in green)
and 6.7 km/s (in red) components at positions A through E in the
B213 Taurus filament (Table 1), for a kinetic temperature of 10 K
and OPR = 0 (see text for discussion). The units of the density are
$10^{4}$ \cc. The blue circles reflect the GBT beam size. The peak
of H$^{13}$CO$^+$ emission is indicated by the white circle, with
diameter reflecting the beam size of the Nobeyama telescope and
center corresponding to core 10b listed in Table 1 of Onishi et al.\
(2002). The dark vectors denote the orientation of the magnetic
field based on optical absorption measurements (Chapman et al.
2011). The underlying image is as in Fig.~\ref{obs} \label{den}}
\end{figure}

Other than the uncertainty in the knowledge of collision rates, the
uncertainty in the derived density is dominated by the uncertainty
in the derived optical depth for the J=2-1 transition. The
uncertainty in the fitted optical depth is affected  by both the RMS
noise, which is 0.09 K and 0.03 K per channel for TMC1C and B213,
respectively, and the line strength. For TMC1C, the sigma for the
derived optical depth of the main hyperfine component is 0.23
compared to its best fit value 0.47. This corresponds to a one sigma
range of J=2-1  corrected integrated intensities between 2.3 K km/s
and 3.0 K km/s. The RMS for the fitted integrated intensities of the
J=10-9 transition is considerably smaller, about 3\%. The range of
line ratios for TMC1C is thus between 0.17 and 0.13. The resulting
uncertainty in density would also depend on the shape of the
excitation curve (Fig.~\ref{rat-all}). The one sigma range of
derived densities for TMC1C are 1.25$\times10^4$ \cc\ to
1.04$\times10^4$ \cc, which corresponds to about a 20\% relative
uncertainty for the derived density for TMC1C. For the  B213
positions, the statistical uncertainties for derived optical depth
are generally on the same order as the value of optical depth
itself.
For position A, the range of derived densities is 1.9$\times$10$^4$ \cc\ to 2.9$\times10^4$ \cc, which corresponds to about a $\pm$20\% uncertainty.
This is representative of the statistical uncertainty in the density  determined  for B213.

In summary, the largest source of uncertainty in the derived density
arises from the imprecisely known kinetic temperature, followed by
the uncertainty in the \h2\ ortho to para ratio. The uncertainty
caused by noise in the line measurement is in the range 20 to 40
percent. Given the reasonable assumptions that have been made for
each of the items as discussed above, we feel that we have a
determination of the volume density through the measurements of
\hc3n\ line ratio that is accurate to within a factor of $\pm$50\%
(factor of three).

The density we derive (next to last column of Table 2) for TMC1C is about 1.1 $\times$ 10$^4$ \cc.
For B213, the derived densities are a factor of  1.4 to 2 higher due to the larger observed line ratio, except for position E,
which has slightly lower density than TMC1C.
The derived densities for the five spatial positions are indicated in Fig.~\ref{den}.

We also derive the column density of \hc3n\ using the observed intensity and the  excitation conditions derived above.
For an optically thin transition, the column density of the upper level is simply proportional to its integrated intensity
\begin{equation}
        N_u = \frac{8 \pi k \nu^2}{hc^3 A_{ul}} \int T_{mb} dv \lc
\label{nu}
\end{equation}
where $T_{mb}$ is the main beam antenna temperature as described in
the data section. We use the integrated intensities of the optically
thin J=10-9 transition to obtain $N(J=10)$. To obtain the total
column density of \hc3n, we utilize the level populations provided
by the same Radex calculations that produced  Fig.~\ref{rat-all} for
collisions with ortho-\h2. The resulting column densities are listed
in the last column of Table 2. The \hc3n\ column density of B213 is
within a factor of two of $10^{12}$ \cm2, while that of TMC1C is one
order of magnitude higher. This is consistent with the well-known
fact that carbon chain molecules have significantly enhanced
abundance in TMC1C.

\section{Discussion: The 3D Structure of B213}
\label{3D}
The volume density measurements enable us to examine the 3D structure of the B213 filament.
We first consider the 2D dimension of the B213 filament projected on the sky.
At the distance of Taurus, one minute of arc corresponds to about 0.04 pc.
The shorter dimension of the B213 filament in the neighborhood of position A is about 4 arcminutes or 0.16 pc as seen in 2MASS extinction.
The filament is elongated by a factor of at least 10 in the orthogonal direction.

The median 2MASS extinction for these positions is about 6.3 mag, which corresponds to $N$(H$_2$) = 6 $\times$10$^{21}$ \cm2.
This is close to that found from $^{13}$CO (Pineda et al. 2010) assuming $X$(CO) = 2$\times$10$^{-4}$, indicating that there is relatively little
depletion of CO, which if present would imply an excessively large total column density.
The modest depletion is consistent with the density we derive and a depletion time scale of a few $\times$10$^5$ yr (Bergin \& Tafalla 2007; Pineda et al. 2010),
if the filament age is not much more than that.
The average density found from our HC$_3$N observations is about $(1.8\pm 0.7)\times10^{4} $ \cc.
From the column density and the volume density, the line of sight dimension is found to be 0.12 pc, which is comparable to the shorter dimension of the
filament projected on the sky.
The derived line of sight dimension is one order of magnitude shorter than  the filament's elongated dimension in the plane of the sky,
which effectively rules out B213 being an edge-on sheet.

As shown in Fig.~\ref{obs}, we detected the higher rotational transition (J=10-9) of \hc3n\ in 14 spatial positions, which are
distributed contiguously along the filament. Detection of the lower transition was limited  in only 5 spatial positions not because of insufficient
excitation, but due to the difficulty of detecting the multiply hyperfine-split J = 2-1 transition.
Fig.~\ref{den} shows density measurements at 5 spatial positions.  The two velocity components present  in these positions yield very similar densities.
All measured densities are within a factor of 2.5 of each other over a projected separation of approximately 0.1 pc.
The line widths at all observed positions are small and their peak velocities are similar.
The volume density measured for these two velocity components cannot result from an isolated ``dense core'' only seen high density tracers such as H$^{13}$CO$^+$.  These densities are also unlikely to be the result of only a few very dense cores much smaller than the beam and resulting beam dilution of their emission.  We seem to be probing lines of sight representative of the B213 filament as a whole.  We cannot, however, rule out completely a highly clumpy cloud structure consisting of a large number of very small dense clumps and an extended low density component. Evaluating such substantially more complex structures will require observations at much higher spatial resolution.

If B213 is indeed a self-gravitating cylinder, an extended low density 'halo' resulting simply from the requirement of hydrostatic equilibrium should be present.
Chandrasekharr \& Fermi (1953), Larson (1985), Inutsuka \& Miyama (1992) and others have studied the configuration and evolution of infinite isothermal
self-gravitating cylinders. The radial density profile of
such cylinder turns out to be of the Plummer type for both the stable and collapsing case with a 'flat' (approximately uniform density) central part
and a halo characterized by an effectively rapidly-dropping power-law density $n$ $\sim$ (1/r)$^4$ (see Eq.~1 in Inutsuka \& Miyama 1992).
In such cases, the density measured by \hc3n\ will probe only the central 'flat' part, as the 'halo' cannot excite the higher rotational transitions.
The dimension thus derived cannot rule out the existence of the extended low density halo, but will characterize the central core of the filament,
which could be considered to the represent the radial size of the filament.

Cylindrical, i.e.\ filamentary,  structures are of increasing interest at the present time due to their prevalence in recent surveys of the ISM,
e.g.~Goldsmith et al.~(2008), Andr{\'e} et al.~(2010), and Molinari et al.~(2010).
The filaments found in these surveys differ by orders of magnitude in terms of physical length, mass, and their star formation content.
These results show that, at least in 2D projection, elongated structures are a common feature of the interstellar medium.
Filaments can be produced in numerical simulations of vastly different physical conditions. Filaments show up in a strongly magnetized
(sub-Alfv\'{e}nic) MHD simulation with ambipolar diffusion (Li et al.\ 2012).
Filaments are also seen in simulation of supersonic converging flows without magnetic field (Gong \& Ostriker 2011).
Filamentary structures are not restricted to fluid simulations (with or without magnetic field, with or without dominant turbulence);
filamentary structures are prevalent in cold dark matter cosmological simulations (Springel et al.~2005), where the most important physics are gravity
and structure growth.

To determine the driving factor of structure formation in the star-forming ISM, the measurement of physical parameters in addition to morphology is critical.
The data presented here are a limited first step.
One particular relevant regime of ISM evolution is molecular cloud formation induced by colliding HI streams
(Ballesteros-Paredes et al. 1999).
A flattened structure (sheet) is expected at the shock front (V{\'a}zquez-Semadeni et al.\ 2006).
The subsequent structure growth depends on the relative importance of different unstable modes, which in turn depend on the physical conditions.
In an illustrative discussion by Heitsch et al.\ (2005), the structures that become visible largely reflect the ratio between volume density and
converging velocity.
When the density is high, thermal instability dominates and the dense structures that grow tend to be contained near the shock front, maintaining
a filamentary appearance.
When the ratio of density to colliding velocity decreases, Kelvin-Helmholtz and nonlinear thin sheet instabilities grow faster and tend to produce
structures perpendicular to the elongation of the shock front.

We are not yet at a stage to differentiate among different simulations, for which the methods themselves are evolving rapidly.
It is, however, tantalizing to describe the evolution of ISM leading to star formation as the following sequence: sheet $\rightarrow$
filament $\rightarrow$ cores.
A flattened structure (sheet) can be produced by colliding streams or other shocks as discussed above.
A self-gravitating sheet will preferentially fragment into filaments (Miyama,  Narita  \& Hayashi 1987).
Inutsuka and Miyama (1992) carried out self-similar calculation of isothermal cylinder and derived a critical filament ''line mass", consistent
with that in Ostriker (1964), beyond which the filament is expected to fragment and collapse into cores.
This last stage of filament--to--cores evolution seems to be emerging as a phenomenon often found in recent observational studies, such
as Hacar \& Tafalla (2011) and Schmalzl et al.\ (2010).

In observing multiple transitions of \hc3n\ toward Taurus filaments, we do not see any suggestions of a  sheet extended along the line of sight.
If a sheet had been present, it must have already evolved or dissipated.
The line width and velocity gradient of \hc3n\ gas are both small at around 0.3 km/s.
At the $\sim$ 10 K temperature of  gas in the filament, this is  marginally supersonic.
For a sound speed of 0.127 km/s corresponding to a line width of 0.3 km/s FWHM, the critical line mass is about 7.6 \Ms/pc.
For B213, this corresponds to a column density of N(\h2) = 4$\times$ 10$^{21}$ cm$^{-2}$, somewhat smaller than the observed value, suggesting
that B213 is susceptible to collapse/fragmentation to produce cores.

The line width of \hc3n\ in B213 filament is similar to that of
\c18o\ in L1517 (Hacar \& Tafalla 2011), while larger than that of
\n2h\ that is observed in the same study. Hacar \& Tafalla also
found a subsonic velocity difference among \n2h\ cores  contained in
the L1517 filaments and were able to fit the radial profiles of
\c18o\ emission using isothermal cylinders. These authors postulate
a two step core formation processes including, first, a subsonic
filament, and second, fragmentation--produced cores. The measured
volume density and line of sight dimension of B213 are consistent
with it being a cylinder-like `true' filament in which  \hc3n\
traces the gas in a manner similar to  \c18o\ in  L1517. The
internal motions of the B213 filament  are still slightly
supersonic, which could mean that B213 is at an earlier evolutionary
stage than L1517, having yet to go through the stage in which the
line width drops and core formation begins.

The volume density measurements and the velocity information presented in this paper will be of additional value when  radiative transfer is
added to various simulations and modeled spectral line profiles can be compared directly with those observed.
The density measurements in this study fit into a developing picture of molecular cloud evolution from turbulent gas to coherent filaments to subsonic cores.

 \section{Conclusions}
 \label{conclusions}
We have determined the volume \h2\ densities along five lines of sight (which include a total 6 distinct components) in
a prominent filament in Taurus, B213, to be on average $(1.8\pm 0.7)\times10^{4} $ \cc.
The density determinations are based on excitation calculations that fit the line ratio between \hc3n\ 10-9 and \hc3n\ 2-1 transitions.
The clearly--resolved hyperfine components of the 2-1 transition enable us to determine the peak opacity of the 2-1 line to be smaller than 1.3.
The line of sight dimension of the dense portion of B213 filament is $\simeq$ 0.25 pc, comparable to the smaller plane--of--the--sky dimension of
the filament, which is about 0.16 pc.
These dimensions are both much smaller than the longer dimension of B213, which is a few pc.
The FWHM line wdith of \hc3n\ is about 0.3 km/s, which is slightly supersonic.
We conclude that B213 is likely a true filament, but that it has not yet formed dense cores throughout.

This work has been partly supported by  China Ministry of Science and Technology under State Key Development Program for Basic Research (2012CB821800).
We thank the staffs at the Green Bank Telescope and the Arizona Radio Observatory for their support during these observations.
We also appreciate very helpful discussions with Steve Stahler on molecular cloud evolution.
We acknowledge  valuable input from Hao Gong and Pak-Shing Li regarding numerical simulations.
We thank Laurent Wiesenfeld for communicating his most recent calculations of HC$_3$N collisional excitation rates.
This research was carried out in part at the Jet Propulsion Laboratory, California Institute of Technology which is supported by the National
Aeronautics and Space Administration.


 \begin{deluxetable}{lccccccccccccc}
 \tabletypesize{\scriptsize}
 \tablewidth{0pt}
 \setlength{\tabcolsep}{0.04in}
 \tablecolumns{14}
 \tablecaption{Fitted HC$_3$N Line Characteristics}
\tablehead{
\colhead{Source} & \colhead{Comp.}& \multicolumn{5}{c}{HC$_3$N 2-1}  &\multicolumn{4}{c}{HC$_3$N 10-9} & \colhead{Line Ratio} & n(H$_2$)  & N(\hc3n)\\
\cline{3-7} \cline{8-11}  \\
&&\colhead{V$_0$} & \colhead{FWHM} & \colhead{T$_{main}$}  &\colhead{$\int T dv$} & \colhead{$\tau_{main}$}   & \colhead{V$_0$} & \colhead{FWHM} & \colhead{T$_{main}$} &\colhead{$\int T dv$}&  \colhead{ \Large{$\frac{\int T(10-9)dv}{\int T(2-1)dv}$} }    & &\\
&&\colhead{km/s} & \colhead{km/s} & \colhead{K}  & \colhead{K km/s}
&  &\colhead{km/s} &\colhead{km/s} &\colhead{K} &\colhead{K km/s} &
& 10$^4$ cm$^3$ & 10$^{12}$ cm$^{-2}$ \\} \startdata
TMC1C &s&     5.4&    0.21&  5.34&     2.6&  0.47&     5.2&    0.29&  1.23&    0.37&    0.15&     1.1&     18.\\
B213 A&b&     5.7&    0.14&  0.45&    0.15&  0.48&     5.6&    0.28&  0.20&   0.058&    0.39&     2.3&     1.5\\
B213 A&r&     6.8&    0.17&  0.22&   0.083&  0.00&     6.7&    0.18&  0.16&   0.030&    0.36&     2.2&    0.80\\
B213 B&b&     5.7&    0.14&  0.14&   0.046&  0.00&     NA&     NA&   NA&     NA&     NA& NA&     NA\\
B213 B&r&     6.8&    0.17&  0.20&   0.076&  0.00&     6.7&    0.19&  0.14&   0.029&    0.38&     2.3&    0.74\\
B213 C&s&     5.6&    0.17&  0.37&    0.14&  0.27&     5.5&    0.29&  0.10&   0.031&    0.22&     1.5&     1.1\\
B213 D&s&     5.6&    0.19&  0.24&    0.10&  0.14&     5.5&    0.19&  0.12&   0.024&    0.23&     1.6&    0.85\\
B213 E&s&     5.6&    0.17&  1.03&    0.39&  1.30&     5.5&    0.18&  0.19&   0.038&   0.097&    0.94&     1.9\\
\enddata
\tablecomments{The units for the fitted FWHM linewidth and central
velocity V$_0$ are km/s. The units for T$_{main}$, the main beam
antenna temperature from the fitting procedures described in the
text, are Kelvin. In the second column, s denotes a single velocity
component, b and r denote blue and red velocity components,
respectively. Column 7 gives the peak optical depth of the F=3--2
(main) HFS component of the J=2-1 transition.   The integrated main
beam temperatures having units of K km/s have been corrected for the
optical depth where appropriate. An optical depth of zero indicates
an optically thin line. ''NA'' indicates  a nondetection.}
\end{deluxetable}

\end{document}